\documentclass[11pt,a4paper]{article}
\usepackage{times}
 \usepackage{graphicx}
 \usepackage{makeidx}
 \usepackage{amsmath}
 \usepackage{amssymb}
 \usepackage[mathscr]{eucal}

 \def \bfd{{\bf d}}

 \def \bfx{{\bf x}}
 
 \def \bfA{{\bf A}}
 \def \bfC{{\bf C}}

 \def \calN{{\cal N}}

\def\bfa{{\bf a}}

\def\bfd{{\bf d}}

\def\bfm{{\bf m}}

\def\bfx{{\bf x}}

\def\bfA{{\bf A}}

\def\bfC{{\bf C}}

\def\bfI{{\bf I}}

\def\calX{{\cal X}}

\def\({\left(}
\def\){\right)}

\def\begeq{\begin{equation}}
\def\endeq{\end{equation}}
\def\begeqarray{\begin{eqnarray}}
\def\endeqarray{\end{eqnarray}}

\usepackage{amssymb} 
\usepackage{amsmath} 

\begin{document}
\title{\bf Informed Proposal Monte Carlo}

\author{Sarouyeh Khoshkholgh \\ Andrea Zunino \\Klaus Mosegaard}

\date{ }
\maketitle

\begin{center}
{Physics of Ice, Climate and Earth\\ 
Niels Bohr Institute, Tagensvej 16\\ 2200 Copenhagen N, Denmark} 
\vspace*{10mm}
\end{center}

\section{Abstract}
Any search or sampling algorithm for solution of inverse problems needs guidance to be efficient. Many algorithms collect and apply information about the problem on the fly, and much improvement has been made in this way. However, as a consequence of the the No-Free-Lunch Theorem, the only way we can ensure a significantly better performance of search and sampling algorithms is to build in as much information about the problem as possible. In the special case of Markov Chain Monte Carlo sampling (MCMC) we review how this is done through the choice of proposal distribution, and we show how this way of adding more information about the problem can be made particularly efficient when based on an approximate physics model of the problem. A highly nonlinear inverse scattering problem with a high-dimensional model space serves as an illustration of the gain of efficiency through this approach.  

%
%
%
%
%

\vspace*{10mm}\noindent
{\small {\it Keywords}: Inverse Problems, Seismic Inversion, Probabilistic Inversion, Markov Chain Monte Carlo, Sampling Methods.}

\section{Introduction}
Over the last 25 years, Monte Carlo methods have been established as a main tool for providing solutions and uncertainty estimates for small- to intermediate-scale, highly nonlinear inverse problems. This development is closely connected to the dramatic increase in computational speed over the last few decades. However, there has also been an increasing demand for solving inverse problems on a larger scale, with more time-consuming forward calculations, e.g., \cite{Fichtner18}, and more complex a priori information, e.g., \cite{Lange12,Grana10}. In this connection it has become clear that straightforward use of standard 
Monte Carlo algorithms is unfeasable, and recent years have seen a surge of modified samplers with more and more sophisticated sampling strategies \cite{Tierney99,Haario06,Vrugt16,Ying20}. Useful improvements have been found, but there is a growing impression amongst applicants that Monte Carlo strategies are fundamentally slow, and that alternatives should be found. This experience has indeed led to improvements where quite efficient solutions, all taylored to the problem at hand through a priori constraints and/or well-chosen simplifying assumptions, have shown promising results (see, e.g., \cite{Fjeldstad18}). Another recent development is an attempt to perform often time-consuming likelihood calculations with neural networks, trained on a very large number of model-data pairs sampled from an a prior probability distribution \cite{Andrieu03,Scheidt18,Nawaz19,Holm-Jensen20}. 

Research in Monte Carlo methods has often been based on a search for new -- often surprising -- inspiration that will allow efficient calculation with simple operations. In the early years of Monte Carlo developments there were many examples of this: Simulated Annealing \cite{Kirkpatrick83}, Hamiltonian Monte Carlo \cite{Duane87}, Simulated Tempering \cite{Marinari92}, Evolutionary Algorithms \cite{Holland92}, etc., all using ideas from other scientific fields to improve sampling, and the benefit has been new ways of building useful intuition to improve our understanding of sampling processes. In recent years we see a continuation of this trend in statistics literature \cite{Roberts09}, and all these methods have brought some success, the degree of which depends on the category of problems they are applied to. 

The 'race of Monte Carlo ideas' has been accompanied by intense discussions in the research community about the efficiency of algorithms. Not only have intuitive ideas been held up against each other, but arguments for and against methodologies have also been accompanied by numerical experiments to support the conclusions. This approach rests apparently on a sound basis, but if we take a closer look at the way algorithm comparisons are typically carried out, we discover a common deficiency: In very few cases, if any, algorithms are compared by solving {\em exactly} the same problem. At the surface, test problems look similar, but a closer look reveals that the information available to algorithms in the same test differs significantly. As a result, comparisons often become meaningless, but there is one thing that seems clear from most comparative studies: The more information about the inverse problem we build into the code of an algorithm, the more efficient the algorithm is. 

The purpose of this paper is to explore how additional information in Monte Carlo sampling may significantly reduce the computational workload. We will first discuss the reasons for the often excessive time-consumption of Monte Carlo strategies. We will then turn to the problem of finding and applying supplementary information to speed up calculations, not from external, independent sources (a priori information), but from the physical problem itself. Our aim will be to apply this information in a way that will not bias the sampling assymptotically. We shall explore and support our findings through numerical experiments.

Our test example will be the acoustic inverse scattering problem for a vertical plane wave hitting a horizontally stratified medium with varying acoustic impedance (product of wavespeed and mass density). This problem is highly nonlinear due to internal multiple scattering (eccoes) and attenuation in the medium. Since our aim is to evaluate solutions and their uncertainties, we use Markov Chain Monte Carlo (MCMC) for the analysis. We compare a straightforward MCMC sampling approach, where the proposal distribution is arbitrary, with one where the proposal mechanism is designed from an approximation to the forward relation. The result is a significant improvement in the algorithm's efficiency. 

\section{Markov Chain Monte Carlo and the Proposal Problem}

\subsection{Proposal Distributions}
The basic idea behind any implementation of Markov Chain Monte Carlo (MCMC) is an interplay between {\em proposals} and {\em rejections}. In each iteration, sampling from a probability density $f(\bfx)$ over a space $\calX$ proceeds from a current value $\bfx$ by first randomly proposing a new value ${\bfx}'$ according to the so-called {\em proposal distribution} $q({\bfx}' | \bfx)$, followed by a random decision where ${\bfx}'$ is accepted, with probability
\begin{equation}
P_{\rm acc}^{{\bfx} \rightarrow {\bfx}'}={\rm min} \( \frac{f({\bfx}') q(\bfx|{\bfx}')}{f(\bfx) q({\bfx}'|\bfx)},1 \).
\label{eq: acc-prob}
\end{equation}
This acceptance probability ensures that, once an equilibrium sampling distribution is established, it will be maintained through {\em microscopic reversibility}, because the probability $P_{\rm acc}^{{\bfx} \rightarrow {\bfx}'} q({\bfx}'|\bfx) f(\bfx)$ of a transition from $\bfx$ to ${\bfx}'$ equals the probability of the reverse transition,
$P_{\rm acc}^{{\bfx}' \rightarrow {\bfx}} q(\bfx|{\bfx}') f({\bfx}')$ \cite{Mosegaard02}. At this point it is important to note that the proposal distribution has no influence on the distribution to which the sampling converges, it only influences the speed of convergence. 

\bigskip\noindent
The two most common types of proposal distributions are:
\begin{enumerate}
\item {\em Local} proposal distributions $q$, where $q({\bfx}'|\bfx)$ depends on the starting point ${\bfx}$. A frequent assumption is translation invariance where $q({\bfx}'|\bfx) = q({\bfx}'  + \bfa|{\bfx}  + \bfa)$ for any shift $\bfa$ in the parameter space. Another common assumption is symmetry : $q({\bfx}' | \bfx) = q(\bfx | {\bfx}')$, and in this case we get a simpler expression expression for the acceptance probability (\ref{eq: acc-prob}):
\begin{equation}
P_{\rm acc}^{{\bfx} \rightarrow {\bfx}'}={\rm min} \( \frac{f({\bfx}')}{f(\bfx)},1 \).
\label{eq_accept}
\end{equation}

\item {\em Global} proposal distributions $q$ that are independent of the starting point ${\bfx}$. This means that $q(\bfx|{\bfx}') = h(\bfx)$ where $h(\bfx)$ is fixed during the sampling process. If $h(\bfx)$ is in some sense close to the target distribution $f(\bfx)$, $h$ is often called a "surrogate" (for $f$).
\end{enumerate}
An MCMC sampler is only efficient if large enough steps (connecting any two areas of high values of $f(\bfx)$ in a few steps) are frequently accepted. This ability critically depends on $q({\bfx}' | \bfx)$, and requires that $q({\bfx}' | \bfx)$ is (at least) locally similar to $f(\bfx')$. This is revealed by a close look at the expression for the transition probability from $\bfx$ to $\bfx'$:
\begin{equation}
P({\bfx}' |{\bfx})= q({\bfx}'|\bfx) \cdot {\rm min} \( \frac{f({\bfx}') q(\bfx|{\bfx}')}{f(\bfx) q({\bfx}'|\bfx)},1 \) \ ,
\label{eq_accept}
\end{equation}
showing that, for $f({\bfx}') \ge f({\bfx})$ and a large $q(\bfx|{\bfx}')/q({\bfx}'|\bfx)$, the transition ${\bfx} \rightarrow {\bfx}'$ is most likely, but for $f({\bfx}') < f({\bfx})$ it is only likely when
\begin{enumerate}
\item $f({\bfx}')$ and $q({\bfx}'|{\bfx})$ are both large at ${\bfx}'$, and \label{cond-1}
\item $q(\bfx|{\bfx}')/q({\bfx}'|\bfx)$ is large \label{cond-2}
\end{enumerate}
We will now see how implementations of local and global proposals may address these questions.

\subsection{Local proposals}
The use of local proposals is an attempt to satisfy the above two conditions: 
\begin{enumerate}
\item This condition is met by aiming to choose a $q({\bfx}'|{\bfx})$ so narrowly that most of $q$'s support coincides with high values of $f$. The underlying assumption here is that $f$ is somehow smooth in the neighborhood of ${\bfx}$. In the absense of external information about the smoothness of $f$, one must usually resort to experimentation with different widths of $q$.
\item This condition is usually met by using a symmetric $q$: $q({\bfx}'|\bfx) = q({\bfx}|\bfx')$. In this way, the ratio $q(\bfx|{\bfx}')/q({\bfx}'|\bfx)$ is always $1$ (and hence never "small").
\end{enumerate}
Local proposals are widely used, but they have at least two serious drawbacks. Firstly, if they are too narrow, the proposed steps will be so small that the algorithm needs many iterations to traverse the parameters space. As a result, many iterations are required to produce sufficiently many independent samples from the space. Secondly, even a very narrow proposal may not approximate the target distribution $f(\bfx)$ very well.

To investigate and exemplify the latter problem in high-dimension\-al spaces, let us consider the case where the target distribution of ${\bfx}$ is Gaussian with covariance matrix $\bfC$ and mean $\bfx_0$:
$f({\bfx}) = \calN_{\bfx} (\bfx_0,\bfC)$.
Assume for illustration that our proposal distribution is an isotropic Gaussian $q({\bfx}|{\bfx}_q) = \calN_{\bfx} (\bfx_q,\bfC_q)$ with mean ${\bfx}_q$ and covariance matrix $\bfC_q$, 
and that we, in the sampling process, have been fortunate to arrive at point with a high value of $f({\bfx})$, say, for simplicity, at its maximum point ${\bfx}_0$. We can now calculate the expected acceptance probability $P^{{\bfx}_0 \rightarrow {\bfx}}$ proposed in the next step by the algorithm:
\begin{equation}
\begin{split}
E(P^{{\bfx}_0 \rightarrow {\bfx}}) 
&= \int_{\calX} \frac{f({\bfx})}{f({\bfx}_0)} q({\bfx}|{\bfx}_0) d{\bfx} \\
&= \int_{\calX} \frac{\calN_{\bfx} (\bfx_0,\bfC)}{\calN_{\bfx_0} (\bfx_0,\bfC)} \calN_{\bfx} (\bfx_0,\bfC_q) d{\bfx} \\
&= \frac{\calN_{\bfx_0} (\bfx_0,\bfC+\bfC_q)}{\calN_{\bfx_0} (\bfx_0,\bfC)}\int_{\calX} 
\calN_{\bfx} (\bfx_1,\bfC_1) d{\bfx} 
\end{split}
\label{eq: Meanf}
\end{equation}
where
\begin{equation}
\bfx_1 
= (\bfC^{-1} + \bfC_q^{-1})^{-1} (\bfC^{-1} \bfx_0 + \bfC_q^{-1} \bfx_0) = \bfx_0
\end{equation}
and 
\begin{equation}
\bfC_1 = (\bfC^{-1} + \bfC_q^{-1})^{-1} \, .
\end{equation}
Since the last integral in (\ref{eq: Meanf}) is $1$, we have the following expression for the expected acceptance probability:
\begin{equation}
E(P^{{\bfx}_0 \rightarrow {\bfx}}) = \frac{\calN_{\bfx_0} (\bfx_0,\bfC+\bfC_q)}{\calN_{\bfx_0} (\bfx_0,\bfC)} 
= \( \frac{{\rm det}\( 2\pi \bfC \)}{{\rm det}\( 2\pi (\bfC+\bfC_q) \)} \)^{1/2} \, .
\end{equation}
Both $\bfC_q = \sigma_q^2 {\bfI}$ (with $\sigma_q^2 > 0$) and $\bfC$ are diagonal in the frame spanned by $\bfC$'s eigenvectors, and if we assume that the eigenvalues of $\bfC$ are $\sigma_1^2 \ge \dots \ge \sigma_N^2 > 0$, where $N$ is the dimension of $\calX$, the eigenvalues of $\bfC+\bfC_q$ are
$(\sigma_1^2+\sigma_q^2), \dots ,(\sigma_N^2+\sigma_q^2)$. From this we have  
\begin{equation}
E(P^{{\bfx}_0 \rightarrow {\bfx}}) = \prod_{n=1}^N
\( \frac{\sigma_n^2}{\sigma_n^2+\sigma_q^2}
\)^{1/2} \, .
\label{eq-eigenC1}
\end{equation}
From (\ref{eq-eigenC1}) we see that for any non-zero values of $\sigma_n$ and $\sigma_q$ we have 
\begin{equation}
E(P^{{\bfx}_0 \rightarrow {\bfx}}) \rightarrow 0 \quad {\rm for} \quad N \rightarrow  \infty \, .
\end{equation}
expressing the influence from the so-called 'curse of dimensionality' on the sampling process. 

If the proposed steps are kept very short ($\sigma_q$ is small compared to all $\sigma_n$), the decrease of $E(P^{{\bfx}_0 \rightarrow {\bfx}})$ with $N$ is slow. But this situation is of no practical value, because adequate sampling by the algorithm requires that it can traverse high-probability areas of $f({\bfx})$ within a reasonable amount of time. For non-negligible step lengths, the situation is radically different. Indeed, if there exists an integer $K$ and a real constant $k$ such that $\sigma_q > k\sigma_n$ for all $n > K$, then $E(P^{{\bfx}_0 \rightarrow {\bfx}})$ decreases more that exponentially with $N$. In other words, if the distribution $f({\bfx})$ is 'elongated' compared to the proposal $q$, that is, if it is broader than $q$ in only a fixed number $K < N$ of directions/dimensions, the mean number of accepted moves will decrease at least exponentially with the number of dimensions.

As an example, let us consider the case where $\sigma_q^2 = 1$, and $\sigma_n^2 = 1/n$. For $N=2$ this gives an expected acceptance probability of $0.4082$, corresponding to a mean waiting time of about $0.4082^{-1} \approx 2.5$ iterations between accepted moves. For $N=10$ the expectation is $1.5828 \cdot 10^{-4}$, and for $N=100$ it decreases to $1.03 \cdot 10^{-80}$, giving a waiting time of about $3.0 \cdot 10^{62}$ years for 1 Billion iterations per second.

The above analysis is carried out under the favorable assumption that the maximum of $f({\bfx})$ has been located by the algorithm, and does not even consider the serious difficulties faced by the sampling algorithm in the initial search for points with high values of $f({\bfx})$ (the {\em burn-in} phase). Hence, it is clear that the proposal mechanism, as defined by $q$, is the Achilles heel of the standard MCMC approach.

\subsection{Global proposals}
A global proposal $q({\bfx}'|{\bfx})$ is independent of ${\bfx}$ and hence it can be written $q({\bfx}'|{\bfx}) = h({\bfx}')$. The use of global proposals seeks to meet the requirements of (\ref{cond-1}) and (\ref{cond-2}) by choosing $h({\bfx}') \approx f({\bfx}')$, ensuring that
\begin{enumerate}
\item $q$ and $f$ are everywhere similar
\item when $f({\bfx}') \leq f({\bfx})$ the condition $q(\bfx|{\bfx}')/q({\bfx}'|\bfx) \gtrapprox 1$ is always met.
\end{enumerate}
In fact, from (\ref{eq_accept}) it is easily seen that global proposals are ideal if they closely resemble the target distribution. In the ideal case where $h({\bfx}') = f({\bfx}')$, the transition probability is equal to $f({\bfx}')$, and the sampler has no rejected moves. Arbitrarily large steps in the sample space are allowed, and therefore all sample points are statistically independent. 

However, the problem with global proposals is to find them in the first place. There are, in principle, two approaches: 
\begin{enumerate}
\item Using, as proposal, a local approximation $h(\bfx)$ to $f(\bfx)$, estimated/interpolated from already visited sample points in the neighborhood of $\bfx$ \cite{Christen05,Ying20}. This proposal may be consistent with (similar to) $f$ in the neighborhood of existing sample points. \label{local-prop}
\item Using a global approximation $h(\bfx)$ derived from external information about $f(\bfx)$, that is, {\em not} derived from already visited sample points. This proposal should be consistent (similar to) $f$ even far away from existing sample points. \label{global-prop}
\end{enumerate}
In the following we shall show an example of the use of global proposals in inverse problems. Our global proposal will be constructed from external information about the target distribution $f$ using an approximate forward function that is independent of known values of $f$. However, before we proceed, we shall first understand the fundamental advantage of (\ref{global-prop}) over (\ref{local-prop}). To this aim, we shall look into an important theorem, proven in the late 90s, namely the No-Free-Lunch Theorem \cite{Wolpert97}.


\section{No-Free-Lunch Theorems and the importance of information}
We will now make an important distinction between {\em blind algorithms} and {\em informed algorithms}. 
We use the following definitions:

\begin{enumerate}
\item A {\em blind algorithm} is an algorithm whose search or sampling is performed only via an {\em oracle}. An oracle is a function that, when called by the algorithm, is able to evaluate the target distribution $f$ at a given point ${\bfx}$. The oracle is used by the algorithm as a black box: No other properties of $f$ than the corresponding inputs and outputs are used. In computer science, blind algorithms are often called {\em heuristics}. For inversion, there are many well-known examples of blind algorithms in use: Regular MCMC, Simulated Annealing, Genetic Algorithms, Neural Networks, etc.
\item An {\em informed algorithm} is an algorithm that, in addition to an oracle, uses known, {\em external} properties of $f$ to guide/improve the search or sampling. By external properties we mean any information about $f$ that is not given by samples from $f$. Examples of informed algorithms used in geophysical inversion are Hamiltonian Monte Carlo, exploiting that for seismic wave fields adjoint methods can be used to efficiently compute misfit gradients \cite{Fichtner18}, and Discriminative Variational Bayesian inversion exploiting knowledge about the statistics of the unknown model in case it is a Markov Random Field \cite{Nawaz19}.
\end{enumerate}
Based on the No-Free-Lunch Theorem (Wolpert and Macready, 1997), Mosegaard (2010) considered limits for the performance of algorithms designed for solution of inverse problems. The conclusion was that all blind inversion algorithms in finite-dimensional spaces (optimization-based as well as sampling-based) have exactly the same performance, when averaged over all conceivable inverse problems. Only an algorithm that take into account more characteristics of the "forward model" than given by the oracle can ensure performance that is superior to blind inversion algorithms. 


We can draw the conclusion that efficient inversion algorithms are the ones that operate in accordance with specific properties of the problem it is aiming to solve. If the problem is linear with known Gaussian noise statistics and a given Gaussian prior, it can be solved in "one iteration" (applying a closed-form solution formula). If the problem is mildly nonlinear with, e.g., Gaussian noise and Gaussian prior, our knowledge that the posterior probability distribution is unimodal will render the problem solvable in relatively few iterations. For a highly nonlinear problem, the situation is, in principle, the same, except that the term "highly nonlinear" usually signals a lack of knowledge of the shape of the posterior. The posterior may be highly multimodal and possess other pathologies, but we may still have some sparse knowledge about it, for instance that it has a certain smoothness. Irrespective of what we know about the target posterior distribution, we have the option of building this information into the algorithm. If we have plenty of information, we can create an efficient algorithm. If we have sparse information, our algorithm will need more computation time. 

Countless methods use interpolation methods to construct local or global approximations to the posterior and to use them as proposals in the sampling process, e.g., \cite{Christen05,Ginting11,Jin11,Stuart19,Ying20} Laloy et al, 2013; Georgia et al, 2019). These methods are useful and may improve performance, but they still suffer from the limitations set by the No-Free-Lunch Theorem, because they do not bring in additional, external information.

In the following we will suggest an approach that allows us to design more efficient inversion algorithms through incorporation of additional, external information about the target distribution. The approach is general and can be used in deterministic as well as in sampling approaches. In this exposition we will focus on MCMC sampling, and our approach will be to replace a traditional, blind proposal mechanism with one built from a simplified forward model. Being based on approximate physics, the chance of obtaining a good global approximation to the posterior is high.

\section{MCMC with Problem-dependent Proposals}
Let us now consider algorithms that bring in new, external information about the target posterior distribution $f({\bfx})$. An approximation $\tilde{f} ({\bfx}) \approx f({\bfx})$, constructed from a simplified version of the physics behind the correct distribution $f$ will be used as a proposal. This proposal will not only be close to $f$ in the neighborhood of points already visited by the algorithm, it is also expected to work well far away from current samples, because it is guided by the physics of the problem. 


\subsection{Linear, Gaussian Problems}
Sampling of solutions to a linear Gaussian problem through MCMC sampling is straightforward. Since we have an explicit expression for the Gaussian posterior, the distribution itself can be used as an optimal proposal. Samples from an $N$-dimensional standard (isotropic) Gaussian (mean $\bf 0$ and covariance $\bfI$) can be generated with, e.g., the Box-M\"uller method, and the desired samples ${\bfm}$ from a $N$-dimensional multivariate Gaussian with mean ${\bfm}_0$ and covariance $\bfC$ can be calculated as 
${\bfm} ={\bfm}_0 + {\bfA}{\bfm}$, 
where ${\bfA}{\bfA}^T = {\bfC}$. The matrix ${\bfA}$ can be found by, for instance, Cholesky decomposition.

\subsection{Nonlinear Problems}
For nonlinear inverse problems, let us consider the general expression for the joint posterior probability in the formulation of Tarantola and Valette (1982):
\begin{equation}
\sigma(\bfd,\bfm) = \frac{\rho(\bfd,\bfm) \theta(\bfd,\bfm)}{\mu(\bfd,\bfm)}
\end{equation}
where $\bfd$ is data, $\bfm$ is the model parameters, and $\rho(\bfd,\bfm)$ and $\mu(\bfd,\bfm)$ is the prior and the homogeneous probability densities in the joint $(\bfd,\bfm)$-space, respectively. The density $\theta(\bfd,\bfm)$ expresses the "uncertainty of the forward relation" between $\bfm$ and data, $\bfd$. For simplicity, let us assume that the homogeneous probability density $\mu(\bfd,\bfm)$, as well as the marginal prior in the model space $\rho_m (\bfm)$ is constant, which leads us to the following expression for the joint posterior:
\begin{equation}
\sigma(\bfd,\bfm) = \rho(\bfd) \theta(\bfd,\bfm)
\end{equation}
Under the further assumption that the observational data uncertainties are small, compared to the modelization errors, we arrive at the approximation
\begin{equation}
\sigma_m(\bfm) = \sigma(\bfd,\bfm) \approx \theta(\bfd_{obs},\bfm)
\end{equation}
This is a very rough approximation, but it should be remembered that we will not replace the accurate posterior by this expression. The approximation will only be used as a global proposal distribution to speed up the search/sampling from the correct posterior.

The question is now how we can find an acceptable expression for $\theta(\bfd_{obs},\bfm)$. In this paper we will adopt the following simple procedure:
\begin{enumerate}
\item Choose a simplified forward function $\tilde{g}(\bfm)$ expressing much of the essential physics, and at the same time allowing an efficient (but probably inaccurate) inversion. This step can be skipped if a direct way to the following step (without a formal inversion) is available.
\item Find a solution $\tilde{\bfm} = h(\bfd_{obs})$ to the simplified problem with an acceptable datafit.
\item Estimate the modelization error introduced by using $\tilde{g}(\bfm)$ instead of the accurate forward function $g(\bfm)$. This error is quantified by the distribution $\tilde{\theta}(\bfd_{obs},\bfm)$, which is also a rough approximation to the posterior ${\tilde{\sigma}}_m(\bfm)$ computed through $\tilde{g}(\bfm)$. The procedure is:
  \begin{enumerate}
  \item The "true" modelization error is 
        $$\delta{\bfm}_{true} = \tilde{\bfm} - {\bfm}_{true} ,$$ 
        but since ${\bfm}_{true}$ is unknown, we compute instead an approximate modelization error 
        $$\delta{\bfm}_{approx} = \tilde{\bfm} - h(g(\tilde{\bfm})) .$$
        The above formula estimates what the modelization would have been if $\tilde{\bfm}$ had been the true model. 
        In case $\tilde{\bfm}$ is close to ${\bfm}_{true}$, we expect that $\delta{\bfm}_{approx}$ will be close 
        to $\delta{\bfm}_{true}$.
  \item Use $\delta{\bfm}_{approx}$ to construct a reasonable approximation to the modelization error distribution 
        $\tilde{\theta}(\bfd_{obs},\bfm)$, centered at $\tilde{\bfm}$. This can be done by assuming a functional form for 
        $\tilde{\theta}(\bfd_{obs},\bfm)$ and by using the components of $\delta{\bfm}_{approx}$ to obtain the parameters of 
        $\tilde{\theta}(\bfd_{obs},\bfm)$. An example of this can be found in the following section.
  \end{enumerate}
\end{enumerate}

\bigskip\noindent

\section{Numerical Example}
To illustrate the gain of computational efficiency obtained by using an even rough approximation to a high-dimensional target posterior as proposal, we shall look at a 1D inverse scattering problem. The unknown model is a horizontally stratified medium with 1000 homogeneous layers. Figure 1B shows the acoustic impedance as a function of distance from the surface. A plane-wave seismic pulse (modeled as a Ricker wavelet) is injected perpendicularly into the medium at the surface, and the data (backscattered waves from the medium) are recorded at the surface (Figure 1A left). The data are synthetic 1-D full-waveform seismic signals generated by the propagator matrix method, containing all multiple reflections, transmission losses and damping effects, so the inverse problem of recovering the model from the data is highly nonlinear. For comparison, an approximate seismogram, computed by convolution of the reflectivity with the Ricker wavelet, is shown in Figure 1A (middle), together with its error (deviation from the correct seismogram) to the right. Figure 1C shows an approximate solution to the inverse scattering problem in the absence of noise, computed by deconvolution, and converted to impedance through trace integration and addition of the slowly varying trend from Figure 1B. The approximate solution requires very little computation time, but is clearly inaccurate (compare to the "true" model in Figure 1B). The purpose of the study is to show how the approximate result can be used to efficiently produce a more accurate solution with uncertainty estimates using Markov Chain Monte Carlo (MCMC).

\begin{figure}
 \includegraphics[width=5.0in]{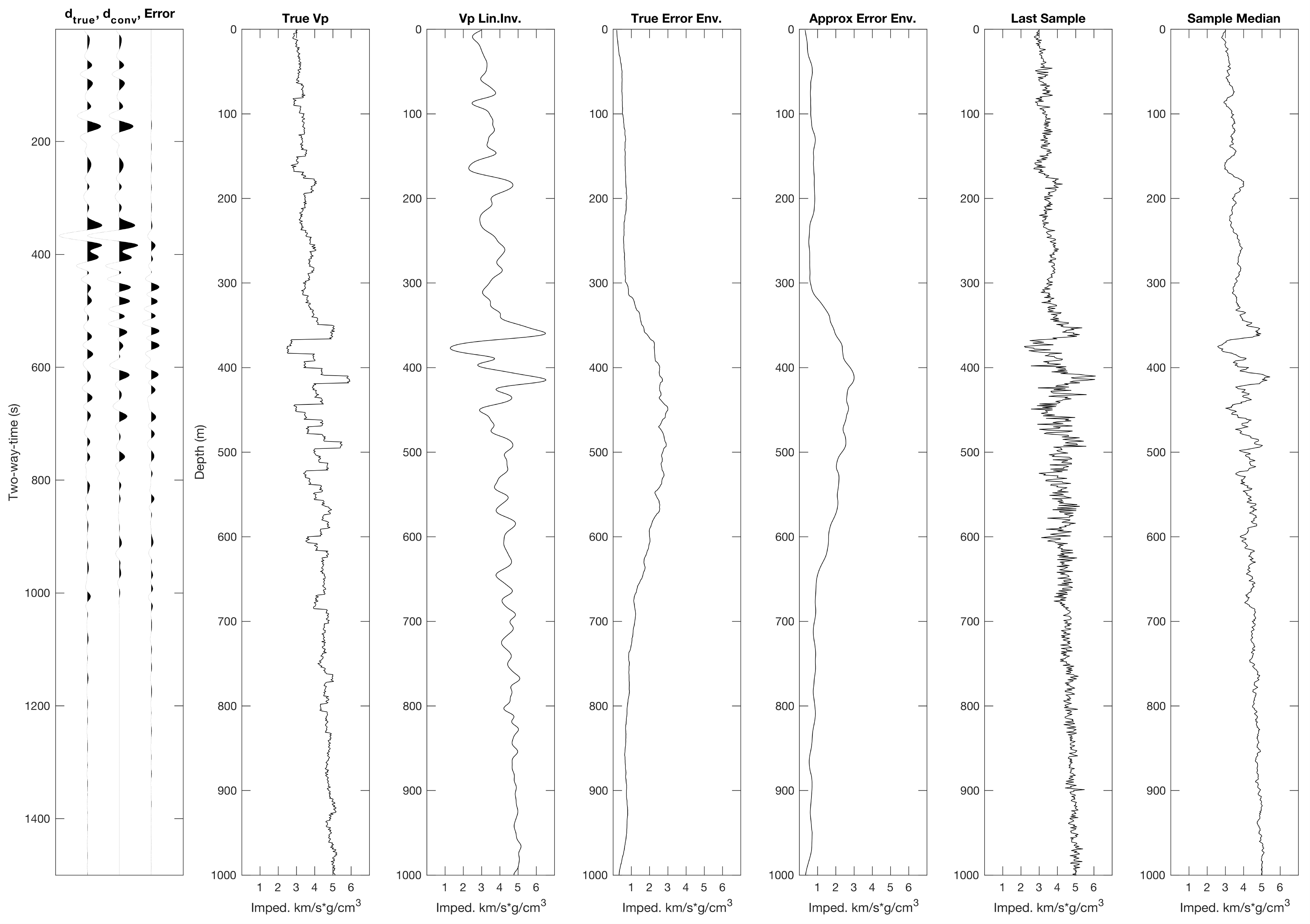}
 \caption{\small (A) Left: Accurate seismogram from B; Center: seismogram computed by convolution; Right: error of the convolution seismogram. (B) True acoustic impedance (C) Acoustic impedance computed by deconvolution (impedance trend from B is added). (D) Envelope of true modelization error (deconvolution impedance minus true impedance). (E) Envelope of estimated modelization error. (F) A sample model from the Informed Proposal Monte Carlo inversion. (G) Median of 10000 sample models.}
 \label{fig: panel}
\end{figure}

\begin{figure}
 \begin{center}
 \includegraphics[width=2.5in]{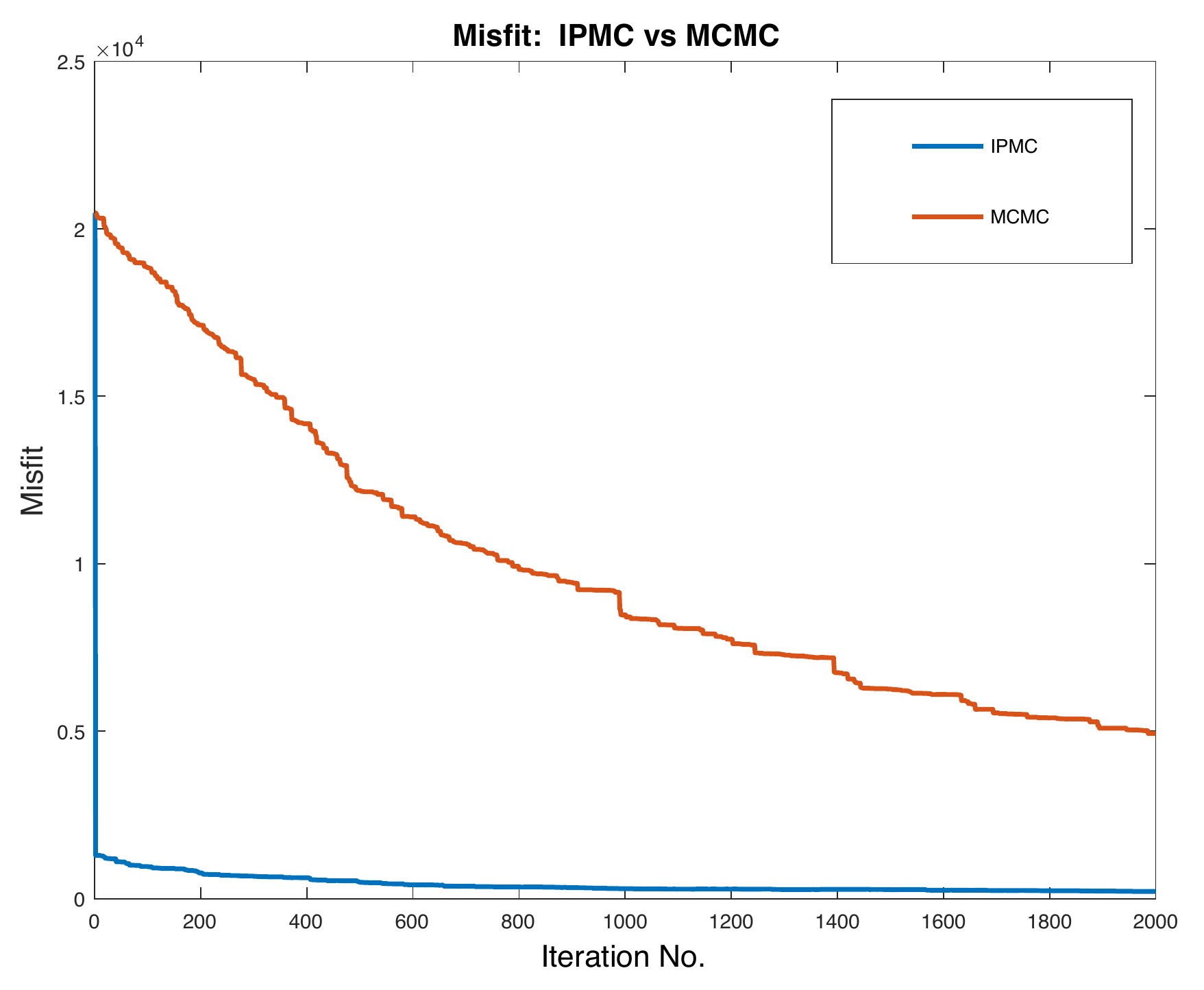}
 \caption{\small Convergence towards equilibrium of a classical MCMC algorithm (upper curve), attempting to sample solutions to our test inverse problem. The lower curve is the fast-converging Informed Proposal Monte Carlo (IPMC) algorithm, which was guided by linearized inversion. In this case the convergence of the guided algorithm was between $10^3$ and $10^4$ times faster than the classical MCMC algorithm (with an tuned, isotropic proposal).} 
 \end{center}
\end{figure}

Our aim is to produce enough samples from the posterior probability distribution in reasonable time, and this raises a well-known problem, namely that the traditional MCMC approach in unfeasible for problems with more than a couple of hundred parameters. Our way of speeding up the sampling is to construct a global proposal distribution for the MCMC sampling using the approximate solution $\tilde{\bfm}$. First, we compute the estimated modelization error vector $\delta{\bfm}_{approx}$ using the method described in the previous section. Figure E shows the envelope of the components of this vector, and for comparison, the true modelization error (known in this synthetic data case) is shown in Figure D. The proposal distribution is then built as a Gaussian with mean $\tilde{\bfm}$ and a diagonal covariance matrix $\bfC_{\theta}$ whose diagonal is the squared components of the envelope function. 

The 1000-parameter problem is now solved in two ways: (1) via a classical MCMC with an isotropic ad-hoc proposal distribution where the step length is adjusted to obtain an acceptance rate of approximately 50\%, and (2) an Informed Proposal Monte Carlo (IPMC) algorithm driven by our proposal derived above.

Figure 2 (upper curve) shows the slow convergence to equilibrium of the classical MCMC in the first 2000 iterations of the inversion process. The lower curve shows the much faster convergence of the algorithm guided by the linearized solution. The improvement in convergence time is significant, in this case between $10^3$ and $10^4$ times faster when started at the model $\tilde{\bfm}$ obtained by linear inversion (deconvolution).

\subsection{Discussion}
It is important to realize that the significantly improved efficiency provided by the physical proposal in this study is {\em not} resulting from prior constraints. Priors generally assign different probabilities to different solutions, but this is not the case with a proposal. A proposal only influences the frequency by which models are presented to the acceptance/rejection algorithm. The bias of the proposal will, asymptotically, be neutralized because it is compensated for in the acceptance probability. In this way it will only influence the efficiency of the sampler, not the asymptotic result. It should, however, be remembered that the most serious problem in non-linear inversion is that the number of models we can practically test is limited. And considering that highly non-linear problems are often so complex that they can only be safely solved with a high number of approximately independent samples from the posterior, it is clear that using an efficient proposal will not only be an improvement in speed, but also a potential improvement in quality of solutions. Simply speaking, we can expect to discover more significantly different solutions (peaks of the target distribution) within the allowed computer resources than with a plain MCMC implementation.

We have illustrated how important it is for the proposal to mimic the posterior in MCMC sampling of solutions to inverse problems. However, the idea of using the physics of the problem to build a posterior-like proposal is not restricted to Monte Carlo sampling. Any method depending on a search for sample solutions or good data fits can potentially benefit from this strategy. In an interesting recent paper on variational full-waveform inversion \cite{Zhang20}, it is shown how variational methods may be used to modify samples from the prior into samples of the posterior in the solution of large-scale inverse problems. It is likely that this class of methods may, in the future, be further improved through application of informed proposal mechanisms.

\subsection{Conclusion}
We have analyzed the impact of proposal distributions on the performance of MCMC sampling methods when applied to the solution of inverse problems. We concluded that the "small step" strategies used in traditional implementations are relatively efficient because they impose a local consistency between the proposal distribution and the target (posterior) distribution: the target probabilities tend to be large where the proposal probabilities are large. Nevertheless, we showed by a simple analytical example that even local consistency may be difficult to obtain when local "small-step" proposals are arbitrary. Furthermore, a main problem with local proposals is the limited step length, which is strongly hampering the exploration of vast, high-dimensional spaces. The volumes of high-probability areas are negligible in such spaces, so burn-in times, and the times needed to pass from one maximum to another can be prohibitive for small-step algorithms. 

Our solution to these problems is to use global proposals built from external information about the target distribution. We propose to use simplified physics of the problem to ensure global consistency between the proposal and the target distribution. The efficiency of this approach will be highly problem-dependent and strongly conditioned on the choice of the external proposal, but we successfully carried out a test on a $1000$-parameter, highly nonlinear inverse scattering problem. Our gain in efficiency was in this case of the order of up to $10^4$.

\subsection{Acknowledgments}
This work was supported by Innovation Fund Denmark through the OPTION Project (5184-00025B). Klaus Mosegaard would like to thank Dr. Amir Khan and colleagues at the Department of Earth Sciences, ETH, for their hospitality and inspiring discussions during the fall 2017 where this work was initiated.

\end{document}